\documentclass[fleqn,usenatbib]{mnras}

\usepackage{newtxtext,newtxmath}
\usepackage{hyperref}
\usepackage{multirow}
\usepackage{tabularx}
\usepackage{xcolor}
\usepackage{bm}
\usepackage[normalem]{ulem}

\usepackage[T1]{fontenc}

\usepackage[utf8]{inputenc} 

\newcommand{\rev}[1]{\textcolor{green}{#1}}
\DeclareRobustCommand{\VAN}[3]{#2}
\let\VANthebibliography\thebibliography
\def\thebibliography{\DeclareRobustCommand{\VAN}[3]{##3}\VANthebibliography}

\usepackage{graphicx}	
\usepackage{amsmath}	
\usepackage{orcidlink}

\newcommand{\stimela}{\textsc{Stimela2}}

\title[TRON IV. M dwarf radio burst]{Mining the time axis with TRON. IV. MeerKAT detects a radio burst from a nearby M~dwarf}

\author[Ramaila et al.]{A.~J.~T.~Ramaila$^{1,2}$\thanks{E-mail: aramaila@sarao.ac.za}, 
O.~M.~Smirnov$^{1,2,3,4,5}$, T.~Myburgh$^{1}$,
D. A. H.~Buckley$^{6,7,8}$, O. D.~Egbo$^{6,7}$,
\newauthor
I. Monageng$^{6,7}$,
A.~Golden$^{9,1}$,
D.~N\'i~Chonchubhair$^{10}$,
X.~Mao$^{11,12}$,
M.~A.~Thompson\orcidlink{0000-0001-5392-909X}$^{13,14}$,
B.~Ngcebetsha$^{2,1}$,
\newauthor
C.~Tasse$^{15,1}$,
I.~Heywood$^{16,17,1,2,4}$, 
J.~S.~Kenyon$^{1}$, 
S.~J.~Perkins$^{2,1}$,
J.~Dawson$^{1}$, 
H.~L.~Bester$^{2,1}$, 
 \newauthor
  J.~S.~Bright$^{4,5}$,
 T.~Blecher$^{2,1}$,
 N.~Oozeer$^{2,1}$,
 V. G. G.~Samboco$^{1}$, 
 I.~Sihlangu$^{2,1}$, C.~Choza$^{4,5}$
\\
$^{1}$Centre for Radio Astronomy Techniques and Technologies, Department of Physics \& Electronics, Rhodes University, Makhanda, 6140, South Africa\\
$^{2}$South African Radio Astronomy Observatory, Cape Town, 7925, South Africa\\
$^{3}$Institute for Radioastronomy, National Institute of Astrophysics (INAF IRA), Via Gobetti 101, 40129 Bologna, Italy \\
$^{4}$Astrophysics, Department of Physics, University of Oxford, Keble Road, Oxford, OX1 3RH, UK\\
$^{5}$Breakthrough Listen, Astrophysics, Department of Physics, The University of Oxford, Keble Road, Oxford, OX1 3RH, UK\\
$^{6}$South African Astronomy Observatory, Cape Town, 7925, South Africa\\
$^{7}$Department of Astronomy, University of Cape Town, Private Bag X3, Rondebosch 7701, South Africa \\
$^{8}$Department of Physics, University of the Free State, PO Box 339, Bloemfontein 9300, South Africa \\
$^{9}$Physics, School of Natural Sciences, Ollscoil na Gaillimhe - University of Galway, University Road, Galway, H91 TK33, Ireland \\
$^{10}$School of Mathematical and Statistical Sciences, Ollscoil na Gaillimhe - University of Galway, University Road, Galway, H91 TK33, Ireland \\
$^{11}$National Astronomical Observatories, Chinese Academy of Sciences, Beijing 100101, People’s Republic of China\\
$^{12}$School of Astronomy and Space Science, University of Chinese Academy of Sciences, Chinese Academy of Sciences, Beijing 100049, People’s Republic of China\\
$^{13}$School of Physics and Astronomy, University of Leeds, Woodhouse Lane, Leeds, LS2 9JT, UK\\
$^{14}$INAF - Osservatorio Astrofisico di Catania, Via Santa Sofia 78, I-95123, Catania, Italy\\
$^{15}$GEPI \& ORN, Observatoire de Paris, Université PSL, CNRS, 5 Place Jules Janssen, 92190 Meudon, France\\
$^{16}$SKA Observatory, Jodrell Bank, Lower Withington, Macclesfield, SK11 9FT, UK\\
$^{17}$Jodrell Bank Centre for Astrophysics, Department of Physics and Astronomy, The University of Manchester, Manchester, M13 9PL, UK
}

\date{Accepted X. Received Y; in original form Z}

\pubyear{2026}

\begin{document}

\label{firstpage}
\pagerange{\pageref{firstpage}--\pageref{lastpage}}
\maketitle 

\begin{abstract}

We report the detection of a minute-timescale, highly circularly polarised ($\sim100\%$) radio burst from an active M dwarf (spectral type M1Ve), PM J14111$-$6155, located 19.6 pc from the Sun. Identified in the 1.28 GHz SMGPS using 8-second integrations from the TRON-pipeline, the burst peaked at 7.7 mJy. Visibility-based dynamic spectra reveal a drifting, band-limited emission feature ($\sim$300 MHz FWHM) with a maximum cut-off near 1.65 GHz, associated with electron cyclotron maser instability (ECMI), implying the existence of large-scale magnetic topologies ($B_* \approx 0.59$ kG). A non-detection ($3\sigma$ upper limit) four hours later points to intermittent or rotationally modulated beamed emission. An analysis of both TESS and ground-based survey data failed to recover an unambiguous rotational period for the optical counterpart. Follow-up SALT spectroscopy confirms its active chromospheric state, exhibiting prominent emission in the Ca {\sc ii} infrared triplet, Balmer lines, and persistent double-peaked H$\alpha$ emission ($44.6\,\mathrm{km\,s^{-1}}$ peak separation) attributable to optically thick self-absorption. This discovery demonstrates that the coordinated use of time-series optical photometry, high-resolution spectroscopy, and high-cadence spectropolarimetric radio imaging enables direct probing of coherent radio emission from the magnetically active atmospheres of nearby M dwarfs.
\end{abstract}


\section{Introduction}

Flaring activity from magnetically active low-mass stars is observed across the electromagnetic spectrum, from radio wavelengths to X-rays \citep{Benz2010}. At radio frequencies, both incoherent gyrosynchrotron emission and coherent bursts are detected, the latter often driven by the electron cyclotron maser instability (ECMI) \citep{Linsky2017, Vedantham2020}. ECMI provides a sensitive probe of energetic particle dynamics and magnetic field strengths in stellar magnetospheres. In density-depleted regions, anisotropic electron populations propagating along converging magnetic field lines can excite the ECMI \citep{Wu1979, Treumann2006}, producing circularly polarised ($\sim 100\%$) radiation at frequencies near the local electron gyrofrequency. Although ECMI emission can occur across a wide frequency range determined by the local magnetic field strength ($f_{\mathrm{cyc}} \approx 2.8\,B\,\mathrm{MHz}$), detecting these beamed, polarised bursts is observationally challenging due to narrow beam geometries and low duty cycles \citep{Hallinan2006,Hallinan2008}.

While periodic coherent emission is frequently observed in ultracool dwarfs \citep[e.g.][]{Rea2026}, serendipitous detections of isolated, highly circularly polarised radio bursts provide vital diagnostics of transient magnetic reconnection and particle acceleration in active stellar coronae \citep{Osten2006}. Because such bursts are governed by narrow beaming geometries and low duty cycles, unravelling their emission mechanisms requires high-cadence spectropolarimetry coupled with optical characterisation. Coherent radio bursts of this type trace strong large-scale magnetic fields and electrodynamic coupling within the system \citep{Zarka1998, Treumann2006, Lilia2026}, providing direct diagnostics of stellar dynamos and magnetic topology in low-mass stars \citep{Morin2010}.

Wide-field high-sensitivity interferometers, such as the South African MeerKAT array \citep{jonas2016}, the Australian SKA Pathfinder (ASKAP; \citealt{Johnston2007}), and the LOw Frequency ARray (LOFAR; \citealt{vanHaarlem2013}), now enable systematic searches for these events directly in the image plane. The TRON pipeline has demonstrated sensitivity to various radio time-domain phenomena, including the recovery of known millisecond pulsars \citep{Smirnov2025a} and additional discoveries \citep{Blecher2026}, as well as distant stellar flare activity in wide-field radio interferometric data \citep{Smirnov2025b}.

Building on this capability, we report the discovery of highly polarised narrowband emission from the nearby active M dwarf PM J14111-6155, identified while reprocessing the SARAO MeerKAT Galactic Plane Survey \citep[SMGPS;][]{Goedhart2024}. Detecting ECMI in this star constrains the emission geometry and duty cycle, probing its active magnetospheric structure \citep{Hess2014}. We combine the radio transient detection from the MeerKAT with follow-up high-resolution optical spectroscopy from the Southern African Large Telescope (SALT), multi-sector Transiting Exoplanet Survey Satellite (TESS) light curves, and archival eROSITA X-ray data \citep{Predehl2021}, providing a comprehensive multiwavelength characterisation of the star's magnetic and coronal activity.

\section{Observations and data reduction}

\subsection{MeerKAT Spectropolarimetric Radio Observations}

The observations reported here were conducted by MeerKAT in conventional synthesis imaging mode, using the L-band (856--1712 MHz) receiver system. The correlator was configured in \rev{`}4k' mode (4096 channels) with an 8-second integration time. SMGPS observations were scheduled as shallow mosaics \citep{Goedhart2024}, with the telescope spending $\le3$ minutes on each target pointing, before slewing to the next target (or, occasionally, to a calibrator), and returning to the same target  $\sim$43 minutes later, repeating the cycle 11 times, using a 9-pointing hexagonal mosaic strategy, resulting in a total observation time of about 7 hours per block.

As highlighted in the first two letters of this series \citep{Smirnov2025a,Smirnov2025b}, TRON inputs are generally obtained by any conventional imaging and calibration workflows. All that is required is a set of calibrated visibilities and a deep sky model. For processing of SMGPS observations, we use the {\tt{CARACal}} pipeline \citep{2020ascl.soft06014J} to perform first- and second-generation calibration, including polarisation calibration, and further develop a custom calibration and imaging pipeline using \stimela ~\citep{stimela2}, specifically tuned to the extended complex emission found in most SMGPS fields. Ramaila et al. (prep) will present a comprehensive description of the SMGPS reprocessing strategy, including improvements in imaging and calibration results. The resulting data products were processed through the TRON pipeline, which performs an automated search for transients in high-temporal-resolution residual images.


\begin{figure}
\centering
   \includegraphics[width=0.95\columnwidth]{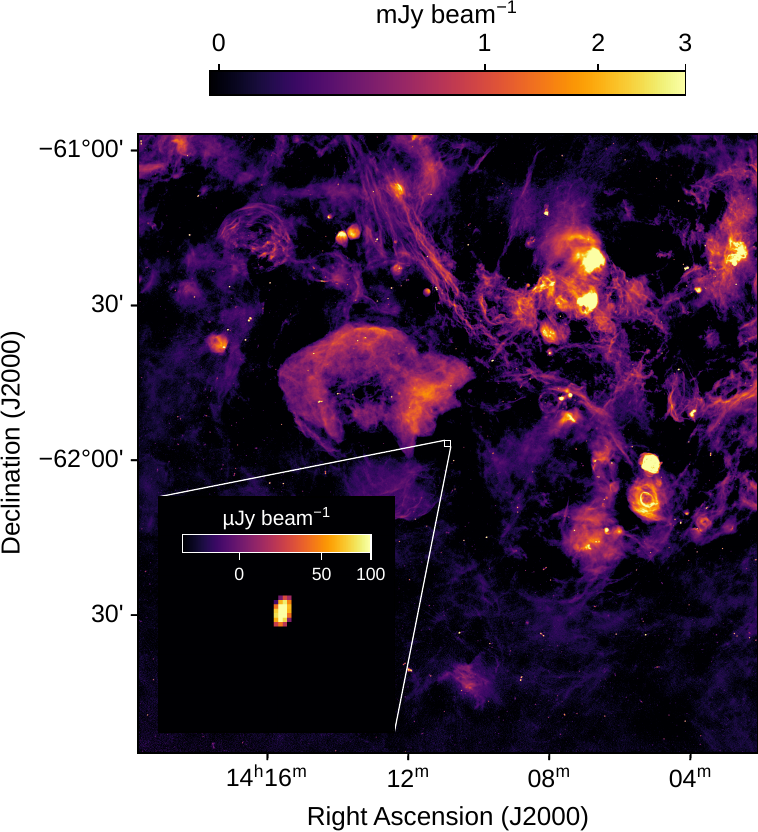}
   \caption{\label{fig:mgpls-big}
   The deep SMGPS MFS mosaic of Tile 16, representing the time-average over 7 hours at an effective frequency of 1.3 GHz. The position of the radio continuum counterpart of PM J14111-6155 is at the centre of the image, within the inset box.}
\end{figure}

\subsection{SALT Optical Spectroscopy}



Optical spectra of PM J14111-6155 were obtained with the Southern African Large Telescope (SALT; \citealt{Buckley2006}) using the High Resolution Spectrograph (HRS) in low-resolution mode ($R \approx 15,000$ in both the blue and red arms). Observations were conducted between 26 May and 10 August 2025 under clear or thin cloud conditions with average seeing of $\sim 1.7\arcsec$. Primary reductions were performed using the {\tt{PySALT}} pipeline \citep{Crawford2016} for overscan, bias, and gain correction. Subsequent echelle extraction and wavelength calibration using ThAr arc spectra and QTH lamp flat-fields were completed with a MIDAS-based package \citep{Kniazev2016}. The resulting spectra cover both blue (3800$-$5550 \AA) and red (5450$-$9000 \AA) regimes. For blue spectra, the Signal-to-Noise ratio (S/N) increased from $\sim 10$ in the bluest orders to $\sim 140$ at 5400 \AA. The red spectra exhibited high S/N, peaking at $\sim 200$ near 8000 \AA. This steep increase in S/N toward longer wavelengths is consistent with the red continuum expected for an active M-type star. 

\begin{figure}
\centering
\includegraphics[width=\columnwidth]{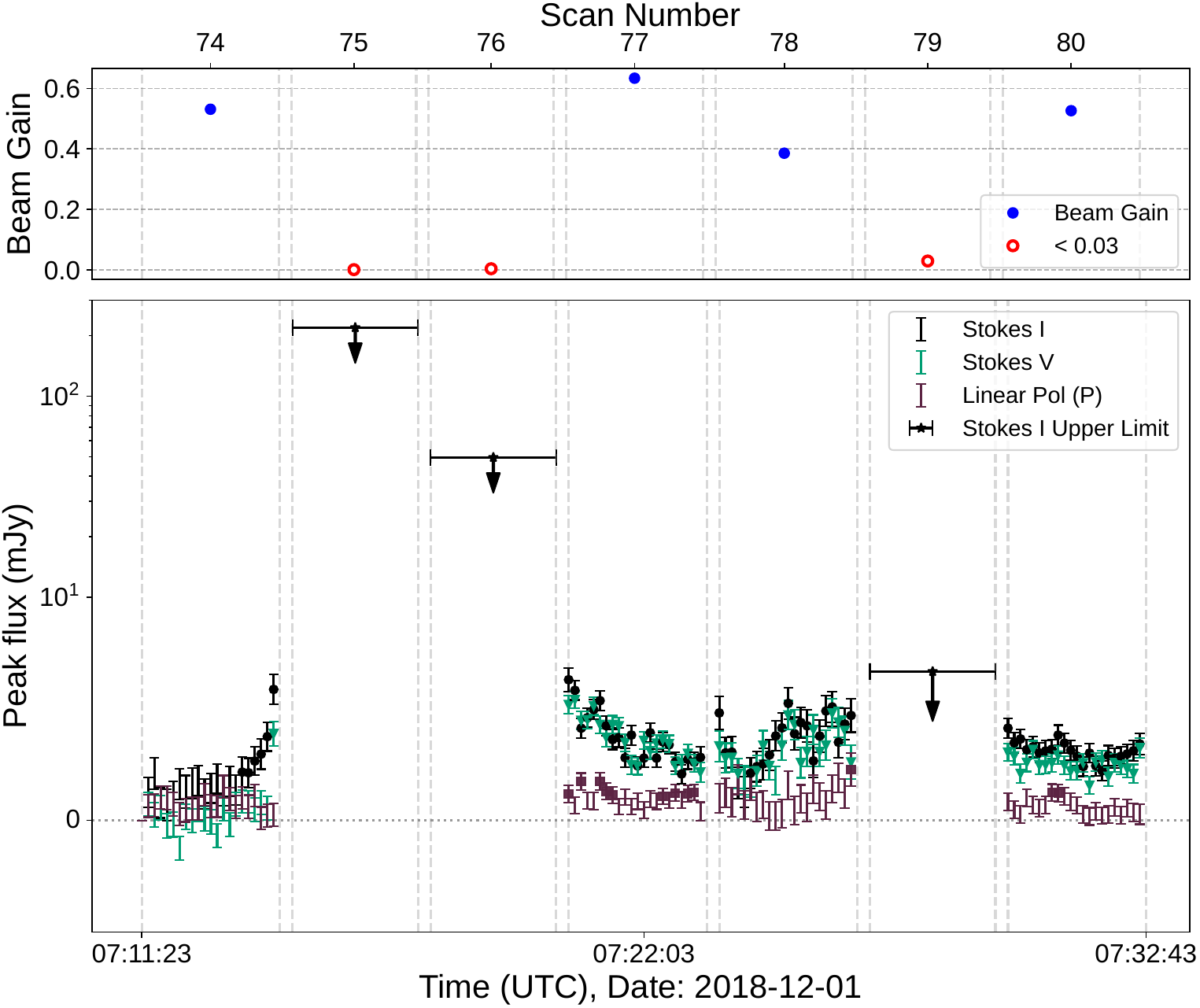}
\caption{\label{fig:curve}Radio light curve of PM J14111-6155 spanning scans 74 to 80, at 8~second integration time. The lower panel shows the flux density measurements at the source position, with detections exceeding $3.5\sigma$ above the local RMS noise indicated by solid circles. The three upper-limit points correspond to mosaic pointings where the source was located well outside the main lobe (primary beam gain $< 3\%$). Dashed points at zero flux represent time slots flagged due to high image RMS, since the telescopes were still settling when they began recording data. The upper panel shows the primary beam gain over time, highlighting variations in instrumental sensitivity due to the target's off-axis position, and their upper-limit flux detection.}
\end{figure}

\begin{figure*}
\begin{tabular}{cc}
\includegraphics[width=0.9\columnwidth]{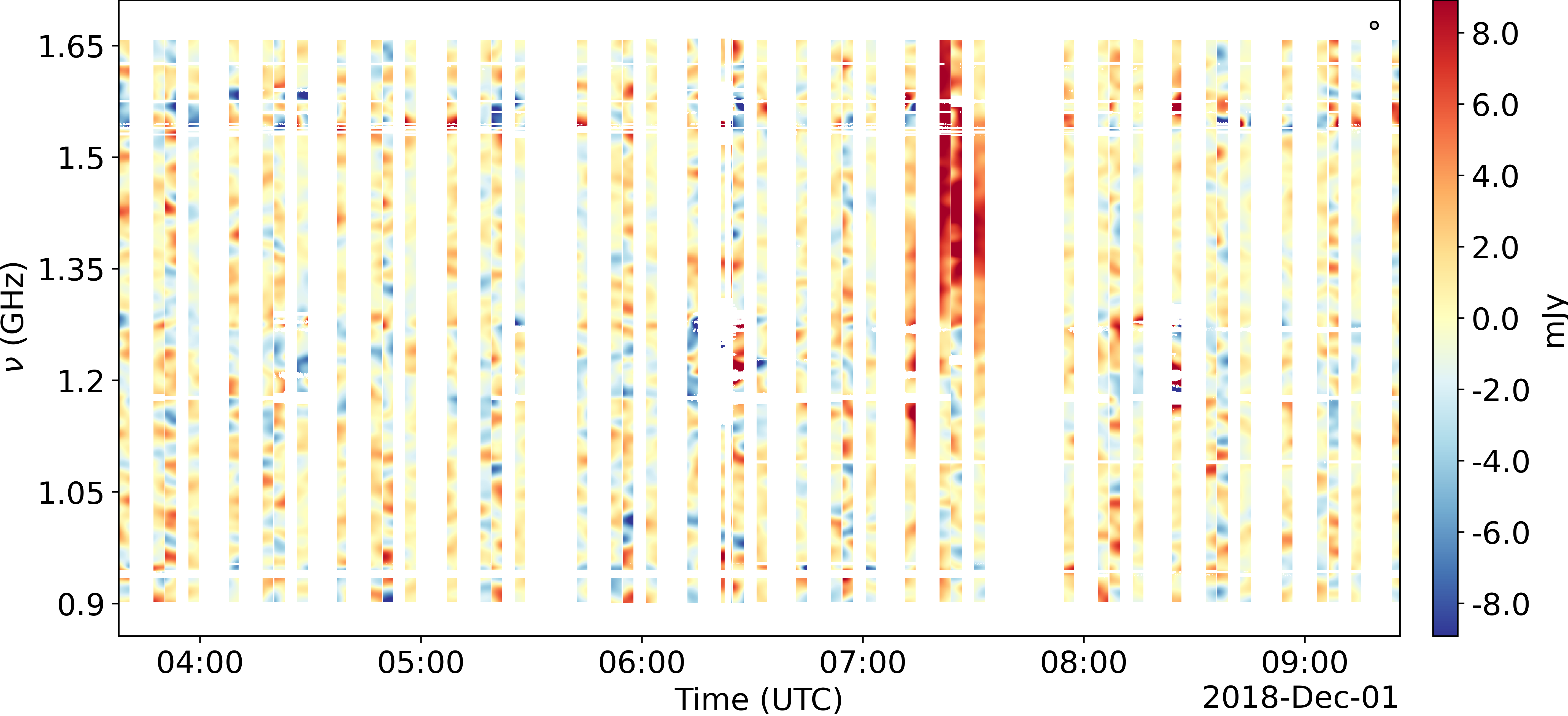} & \includegraphics[width=0.9\columnwidth]{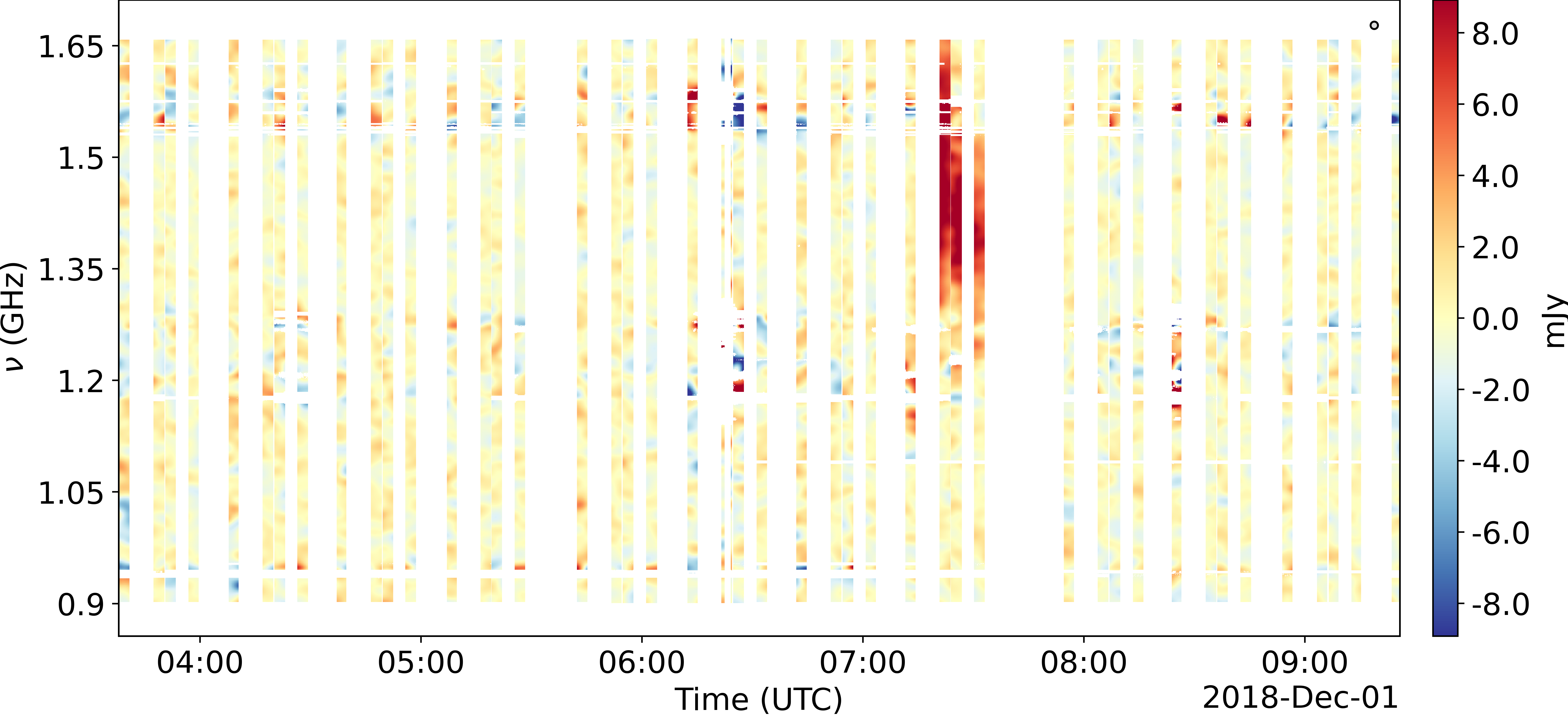} \\
\includegraphics[width=0.9\columnwidth]{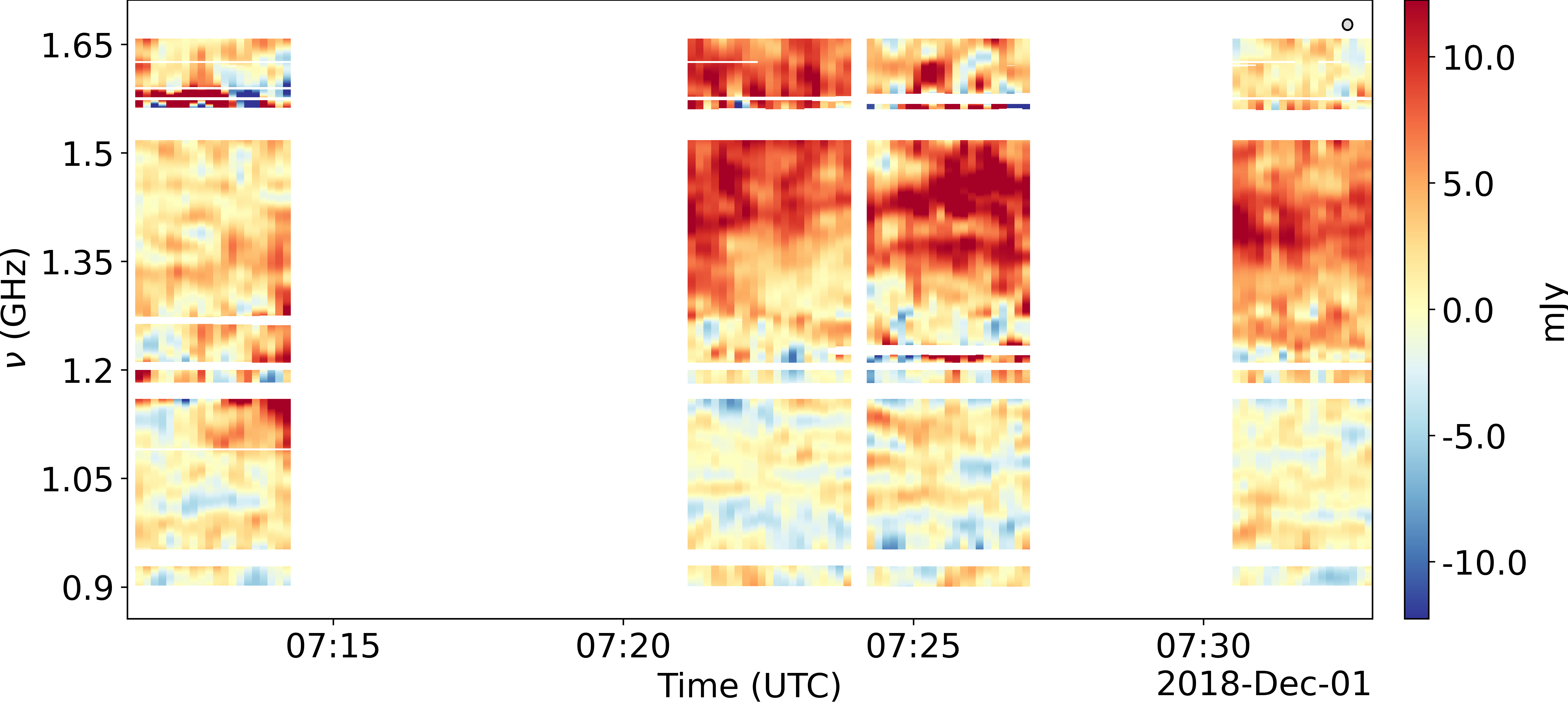} & \includegraphics[width=0.9\columnwidth]{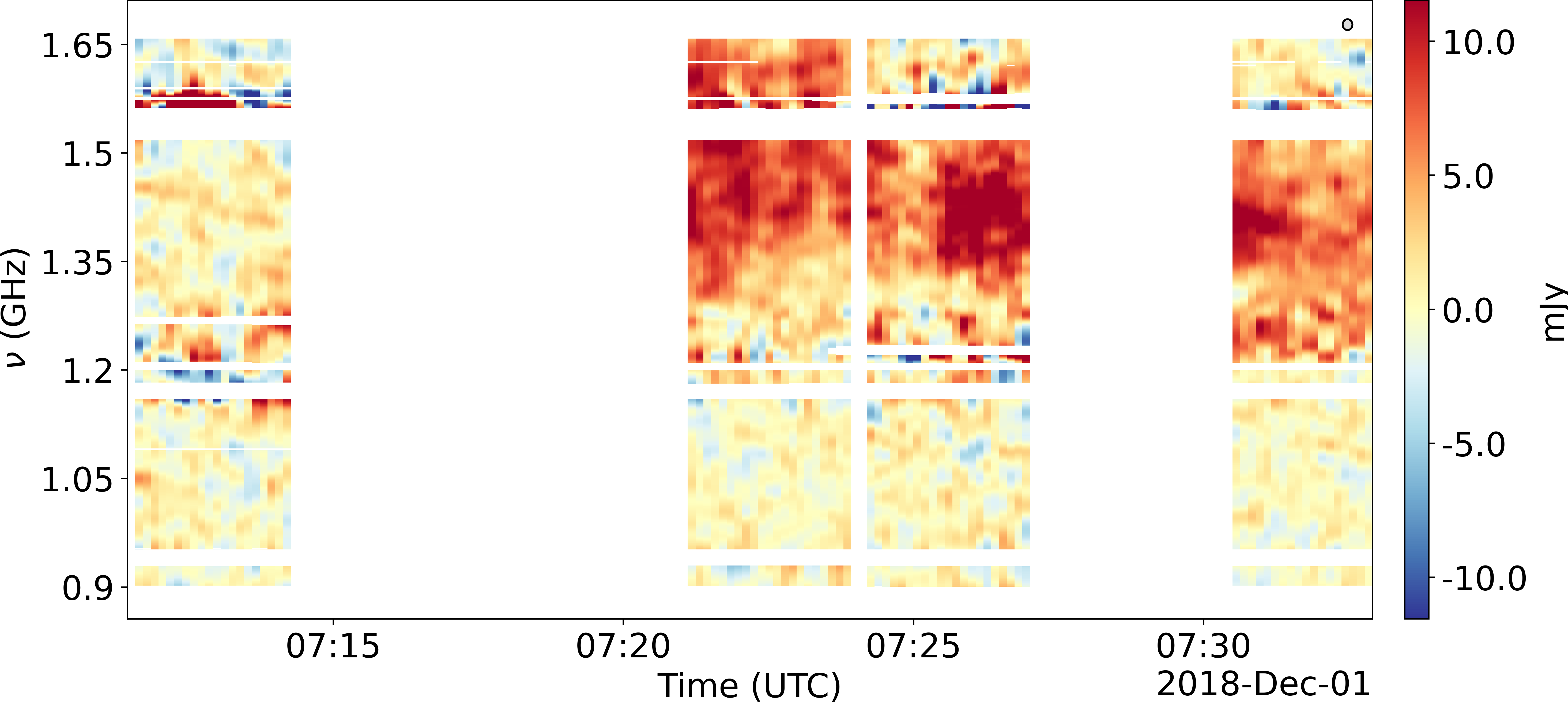} \\
\end{tabular}

\caption{\label{fig:l3-ds} Primary-beam-corrected Stokes $I$ (left) and $V$ (right) dynamic spectra of PM J14111$-$6155. Top panels are smoothed with a Gaussian kernel of $\mathrm{FWHM} = 10\text{ MHz} \times 120\text{ s}$, and bottom panels with a Gaussian kernel of $\mathrm{FWHM} = 15\text{ MHz} \times 10\text{ s}$. The FWHM of the kernel used for smoothing is indicated by a grey ellipse in the upper-right corner of each panel. Bottom panels zoom in on the scans with detected emission, corresponding to the dashed rectangular region in the top panels. Gaps in the dynamic spectra correspond to telescope slewing, calibrator scans, and fully flagged (RFI-contaminated) channels. Note that the noise across time and frequency is non-uniform due to the primary beam correction applied across different pointings.}
\end{figure*}

In order to find any evidence of binarity, longslit low resolution (R $\sim$ 830) spectroscopy of PM J14111-6155 was conducted using the Robert Stobie Spectrograph on SALT \citep[RSS;][]{Burgh2003,Kobulnicky2003,Smith2006}. Ten repeat 200 s exposure spectra were obtained on 2 June 2026 using the low  PG0900 grating at a grating tilt of $12.5^\circ$ with slit width $1.5\arcsec$. The spectrum covered the wavelength range 3200 \AA\ to 6300 \AA\ with a resolution of 5.8 \AA. 

The RSS spectra were reduced using the \texttt{PySALT} package \citep{Crawford2010}, which includes bias subtraction, flat-fielding, amplifier mosaicing, and flagging cosmetic defects. An observation of a spectrophotometric standard star, taken in twilight, was used to determine a relative flux calibration. The spectra were wavelength-calibrated, background-subtracted, flux-calibrated, and extracted using standard IRAF\footnote{IRAF is distributed by the National Optical Astronomy Observatory, which is operated by the Association of Universities for Research in Astronomy under cooperative agreement with the National Science Foundation.} procedures. 
Finally, heliocentric velocity corrections ($-5.39$ to $-5.42$ km~s$^{-1}$; e.g., \citealt{Crawford2016}) were applied across the 17:33:03 to 18:05:34 UTC exposure sequence.

\section{Radio Burst Detection and Analysis}

The radio transient was discovered by high-cadence imaging within the SMGPS using the TRON pipeline. Although identified by its temporal variability, the burst was sufficiently energetic to be detected in the time-averaged multi-frequency synthesis (MFS) mosaic (Fig.~\ref{fig:mgpls-big} highlighted with the inset box). This continuum counterpart, characterised by a flux density of $\sim$0.2 mJy, is listed as a point source in the SMGPS catalogue \citep{2026MNRAS.546f1849M} and identified as a radio star by \cite{Egbo:2025}, and our analysis further characterises its radio transient behaviour. The transient coincides with the high-proper-motion, low-mass star PM J14111-6155. This star is an M dwarf with spectral type M1Ve \citep{torres2006}, at a distance of $19.59 \pm 0.01$ parsecs, based on a \citet{2020yCat.1350....0G} DR3 parallax of $51.04 \pm 0.02$ mas and has a proper motion $(\mu_{\alpha}^*, \mu_{\delta}) = (154.8, -45.1) \text{ mas~yr}^{-1}$ (a $0.55''$ shift at the MeerKAT epoch). It was initially identified as an M dwarf during a search of the local solar neighbourhood \citep{riaz2006}. PM J14111-6155 (also known as TYC 9009$-$33$-$1, UCAC4 141$-$132480, and 2MASS J14110874$-$6155469) is known to be active, with H$\alpha$ EW of 3 \AA\ \citep{riaz2006}, and is associated with a SWIFT X-ray detection \citep{Reynolds2012}.

TRON detected the emission in the 8-second images with a significance of approximately $10\sigma$, reaching a peak flux density of $7.7 \pm 0.1$~mJy (see Fig.~\ref{fig:curve}). At a distance of 19.6~pc, this implies a brightness temperature of $T_\mathrm{b} \gtrsim 10^{12}$~K (assuming a source size comparable to the stellar radius), confirming a coherent emission mechanism. The peak midpoint occurred on 2018 December 1, at 07:14:11 UTC.

To assess the spectral morphology during the brightest detected emission interval (scan 74 at 07:11:23--07:14:19\,UTC in Fig.~\ref{fig:curve}) independently of the dynamic spectra, we performed snapshot imaging across the 856 - 1712\,MHz band divided into two to five sub-bands. Point-source fits to these synthesised snapshot images reveal a peaked spectrum across all frequency subdivisions, reaching a maximum recorded flux density of $15 \pm 2$\,mJy near 1.3\,GHz and declining to $\lesssim 5$\,mJy below 1.1\,GHz and above 1.5\,GHz. Because the source was still brightening when the scan ended, the detected flux density reflects the maximum rather than the true burst peak, whose duration and profile remain unconstrained by telescope slewing gaps.

PM J14111$-$6155 is the third serendipitous image-plane radio transient detected with MeerKAT linked to an M-dwarf star, following similar events reported by \cite{2022MNRAS.510.1083D} and \cite{2022MNRAS.513.3482A}. Unfortunately, the initial detection occurred right at the end of a target scan (scan 74, Fig.~\ref{fig:curve}), and immediately after the detection interval, the telescope was slewed to a new pointing centre (scan 75). With 16\,s lost to slewing, we then have an observation in scan 75 that occurs 24\,s later, with the source considerably further off-axis, and thus with a much lower effective sensitivity (power beam gain across the band is $0.22-5\times10^{-4}$), yielding no detection. The per-integration sensitivity towards the source is an order of magnitude lower (57 mJy intrinsic image RMS at the peak of the spectrum), so we can only say that the burst did not get an order of magnitude brighter.
Target effective sensitivity varies as the telescope moves between pointings (Fig.~\ref{fig:curve}), with low-sensitivity integrations marked red in the upper panel.

The dynamic spectrum of the burst, generated using {\tt{RIMS}}\footnote{\url{https://github.com/saopicc/RIMS}} \citep{Tasse2026} and {\tt{inspect-dynspec}}\footnote{\url{https://github.com/ratt-ru/inspect-dynspec}} are shown in Fig.~\ref{fig:l3-ds}. These reveal a narrowband, highly circularly polarised signal that is strongly time-localised and morphologically consistent with coherent emission processes. To enhance the signal-to-noise ratio, the data are whitened using 30 off-target pointings (within the same MS) and smoothed with a 2D Gaussian kernel \citep{Zhang2025}. Decoupling the off-target statistics preserves the intrinsic morphology, bandwidth, and duration of the emission.

The emission is confined to a bandwidth of approximately $300\text{ MHz}$ (FWHM) centred near $1.3\text{ GHz}$, consistent with the peak detected in the snapshot sub-band imaging. The high-resolution dynamic spectrum (Fig.~\ref{fig:l3-ds}) reveals an emission envelope bounded by distinct upper and lower cut-off trajectories. The upper cut-off frequency exhibits a systematic downward drift, shifting from the upper instrumental band edge at $\nu_{\mathrm{cut, upper}} \approx 1.65\text{ GHz}$ down to 
approximately $1.50\text{ GHz}$ towards the end of the event. Although the initial $1.65\text{ GHz}$ limit coincides with heavy RFI flagging, this downward drift confirms a real astrophysical boundary. Assuming fundamental gyrofrequency emission, this traces a real-time reduction in the maximum local magnetic field strength from $B_{*} \approx 0.59$~kG to $0.54$~kG. Concurrently, the lower cut-off remains bounded near $\nu_{\mathrm{cut, lower}} \approx 1.35$~GHz. This restricted bandwidth indicates emission from a highly localised region with a defined magnetic field gradient ($\Delta B \approx 0.11$ kG). This lower-frequency boundary can be explained by several mechanisms. One possibility is that the absolute lower boundary is physically governed by the local plasma frequency ($\nu_p \propto \sqrt{n_e}$), the critical density threshold where ambient coronal plasma screens escaping maser emission \citep{Zhang2025}. Alternatively, the cutoff may mark the spatial limit where the unstable electron pitch-angle distribution ceases to support maser amplification, or reflect changes in the beaming geometry as the narrow maser emission cone sweeps across our line of sight along the active magnetic field lines \citep{Treumann2006}.

Furthermore, analysing archival x-ray observations from the \textit{eROSITA} telescope \citep{Predehl2021}, PM J14111$-$6155 was detected across 10 scans with an observed unabsorbed flux of $F_{\mathrm{X}} = (4.4 \pm 0.3) \times 10^{-13}\text{ erg s}^{-1}\text{ cm}^{-2}$ (0.2--2.3 keV). At $d = 19.6$ pc, this corresponds to a quiescent soft X-ray luminosity of $L_{\mathrm{X}} \approx 2.0 \times 10^{28}\text{ erg s}^{-1}$. The merged spectrum is well described by an optically thin thermal plasma model (\textsc{apec}; \citealt{Smith2001}) with a coronal temperature of $kT = 0.74^{+0.05}_{-0.07}$ keV. In contrast, the peak spectral radio luminosity of the burst is $L_R \approx 4.6 \times 10^{15} \text{ erg s}^{-1} \text{ Hz}^{-1}$.
On the $L_\mathrm{X}$–$L_\mathrm{R}$ plane, it exhibits a severe radio over-luminosity, lying orders of magnitude above the empirical Güdel-Benz relation \citep{GuedelBenz1993}.

\begin{figure}
\centering
\includegraphics[width=0.95\columnwidth]{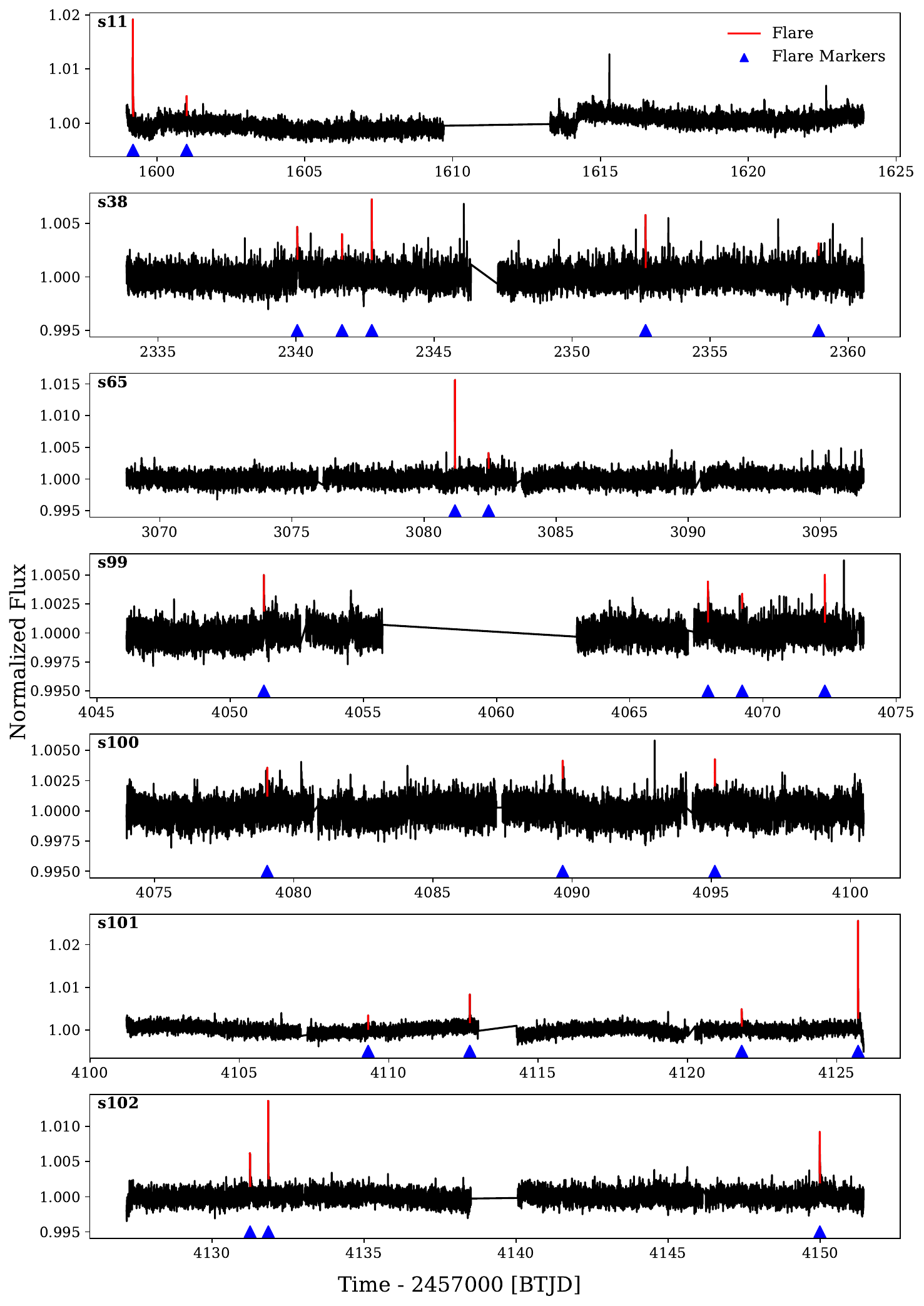}
\caption{\label{fig:l3-tess} 
TESS light curves of TIC 332286707, from top to bottom, Sectors 11, 38, 65, 99, 100, 101 and 102, at 120-second cadence. The relatively quiescent nature of PM J14111-6155 is readily apparent. Flares as detected by {\tt{AltaiPony}} are highlighted in red and marked by blue triangles for visibility.}

\end{figure}

During the burst, the emission exhibits a high degree of circular polarisation (up to $100\%$ right-handed, following the IAU convention), with no significant detection of linear polarisation (yielding a $3\sigma$ upper limit of $p_{\rm linear} < 10\%$ at peak flux). This high circular polarisation fraction, together with the restricted bandwidth and abrupt temporal onset, provides strong evidence for ECMI-driven auroral processes analogous to those observed in ultracool dwarfs and planetary magnetospheres \citep{Treumann2006, Hess2014}. Because ECMI radiation is intrinsically beamed into a thin hollow cone at large angles to the magnetic field vector \citep{Treumann2006}, this emission may either reflect geometric rotational modulation from an oblique rotator \citep{Trigilio2000, Leto2017} or represent an intrinsic, short-lived transient activation of the maser source, such as localised magnetic reconnection or particle acceleration \citep{Hess2014}.  The 20-minute duration, narrow relative bandwidth, severe radio over-luminosity, high brightness temperature ($T_\mathrm{b}$ $> 10^{12} \text{ K}$) and $\sim 100\%$ circular polarisation rule out incoherent gyrosynchrotron emission and favour a coherent mechanism.

The single-epoch detection of PM~J14111$-$6155 and its non-detection $\sim$4\,hours later (undetected below a $3\sigma$ upper limit of $0.3\text{--}0.5\text{ mJy}$ per scan) can be explained by several scenarios. Narrow geometric beaming from an oblique rotator produces low-duty-cycle pulses that readily escape detection with sparse temporal sampling, especially in the absence of an independent rotation period to phase the observations. Alternatively, the burst may represent a transient reconnection event which only temporarily energised the electron population. Dedicated high-cadence radio monitoring is needed to distinguish between transient flaring from stable, rotationally modulated auroral pulses.

\begin{figure}
\centering
\includegraphics[width=\columnwidth]{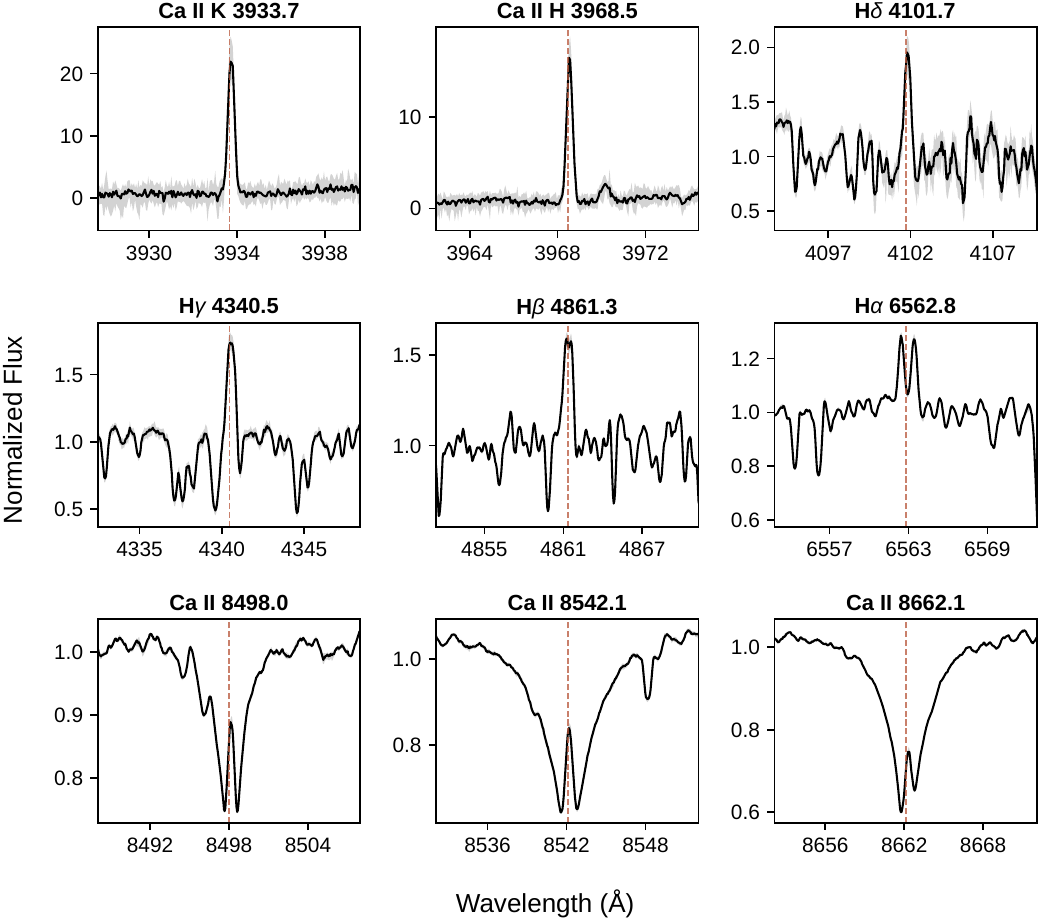}
\vspace{-5mm}
\caption{Chromospheric activity diagnostics in SALT/HRS spectra of PM~J14111$-$6155. Each panel displays the median of 28 spectra across 10 nights (2025 May 26--August 10), with shading denoting the 16th--84th percentile epoch-to-epoch scatter, and dashed lines indicating rest wavelengths. Spectra are transformed to the heliocentric frame and locally normalised. The Ca\,{\sc ii} K (3933.66\,\AA) and H (3968.47\,\AA) lines show narrow emission cores reaching $\sim$22 and $\sim$16 times the near-zero local continuum, with H$\epsilon$ visible at 3970\,\AA. The Balmer lines ($\text{H}\delta$--$\text{H}\alpha$) are all in emission, were $\text{H}\alpha$ shows a double-peak with a separation of $44.6\pm0.6\text{ km s}^{-1}$. The Ca\,{\sc ii} infrared triplet appears in absorption with superimposed core emission reversals.}
\label{fig:l3-chromo}
\end{figure}

\begin{figure}
\centering
\includegraphics[width=\columnwidth]{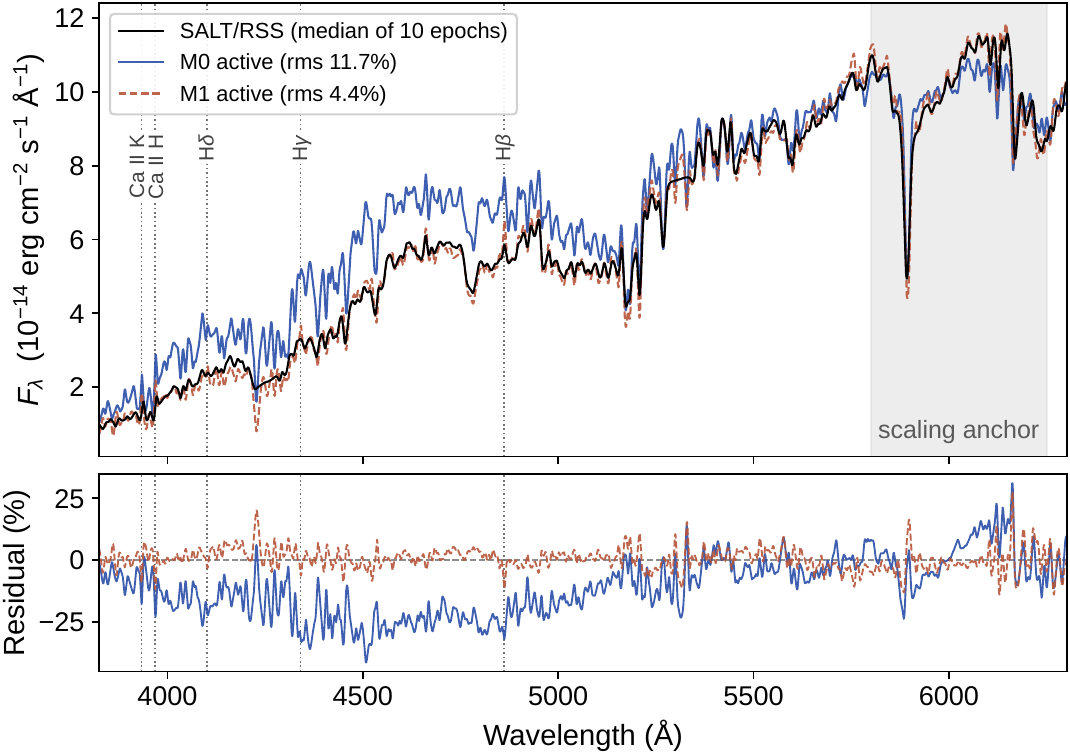}
\vspace{-5mm}
\caption{The flux-calibrated SALT/RSS spectrum of PM\, J14111$-$6155 (solid black line,  median of ten 200\,s exposures from 2026 June 2), compared with the active M0 (solid blue line) and M1 (dashed orange line) spectral templates of \citet{Bochanski2007}. The normalised template spectra are scaled to match the observed flux in the 5800$-$6250\,\AA\ region (grey shaded band). Vertical dotted lines mark the rest wavelengths of Ca\,{\sc ii} H and K, and higher-order Balmer lines, with H$\alpha$ lying redward of the RSS wavelength coverage. \textit{Lower panel:} Residuals after subtracting each scaled template, expressed as a percentage of the median observed flux. Over the 3825$-$6302\,\AA\ range, the M1 template yields an rms of 4.4\%, whereas the M0 template yields 11.7\% and falls 20--40\% below the data between 4000 and 5000\,\AA.}
\label{fig:rss_templates}
\end{figure}

\section{TESS flares, Optical Spectral Diagnostics and Classification}

As PM J14111-6155 is a nearby bright M dwarf, it has been the focus of several studies to select targets for exoplanet detections \citep[e.g.][]{Frith2013,Gaidos2014,Reiners2020,Hardegree-Ullman2023} and, as AP49383418 catalogued in the ASAS-SN archive, showing stochastic variable behaviour consistent with this class of M dwarf with a $m_{V}$ = 10.99 over 170 epochs between March 2016 and September 2018 \citep{2014ApJ...788...48S,2017PASP..129j4502K}.

PM J14111$-$6155 identified as TIC 332286707, has been observed by NASA’s Transiting Exoplanet Survey Satellite (TESS; \citealt{Ricker2015}) on seven separate occasions in 2019, 2021, 2023 and 2026 (TESS sectors 11, 38, 65, 99, 100, 101, 102), at 120\,s cadence, which we adopt throughout. Each sector data set was downloaded and detrended using a Savitzky-Golay filter with the {\tt{LightKurve}} package \citep{2018ascl.soft12013L} and searched for flare candidates using {\tt{AltaiPony}} \citep{2016ApJ...829...23D, 2021AA...645A..42I} (see Fig.~\ref{fig:l3-tess}). Flare candidates are required to show a flux enhancement above the $3\sigma$ level for at least three consecutive points to be considered a flare. In total, 23 flare events were detected across a $\sim$ 189-day baseline, confirming it to be magnetically active but by no means significantly so.

TESS observations indicate that only $\sim$10\% of M0-M3 dwarfs exhibit such activity \citep{Gunther2020}. Active dwarfs in this regime typically produce $10^{-2}$ to 1 flare per day, dominated by low-energy ($< 10^{31}$ erg) events. While PM J14111-6155 shows activity signatures (H$\alpha$ and X-ray emission), its sparse TESS flare activity detections remain consistent with the flare activity expected for its spectral class. We derive a flare rate with a Poisson event-counting uncertainty of  $0.139 \pm 0.029$ per day. This is higher than the flare rates typically reported for inactive early M0-M1 dwarfs, but is consistent with the flare frequencies observed in magnetically active M0-M1 dwarfs in Kepler and TESS surveys \citep[e.g.,][]{Kowalski2009, Hilton2010, Davenport2016, Raetz2020}. Each separate TESS sector time series was analysed for evidence of periodicity using a Lomb-Scargle periodogram, as was the combined data set, yielding no consistently significant signal. This was also the case following a similar analysis of AP49383418 in the ASAS-SN archive. We note that TIC 332286707 was one of 7,481,412 stars studied as part of the recent TESS All-Sky Rotation Survey \citep{2026arXiv260305586B} with no evidence of a periodic signal.

Moreover, the \ion{Ca}{ii} infrared triplet lines ($\lambda8498, 8542, 8662$) in our high-resolution SALT/HRS spectra display broad photospheric absorption wings with prominent emission reversals in their cores, confirming the high surface gravity of a main-sequence dwarf \citep{2023A&A...671A.162M}. Furthermore, prominent emission is present across all optical line diagnostics: \ion{Ca}{ii}~H and K exhibit intense, narrow emission cores reaching $\sim$16 and $\sim$22 times the local continuum, with H$\epsilon$ ($\lambda3970.1$) visible in emission on the red wing of \ion{Ca}{ii}~H.
The Balmer series (H$\delta$, H$\gamma$, H$\beta$, and H$\alpha$) is observed in emission, with H$\alpha$ having a double-peaked profile with a peak-to-peak velocity separation of $44.6 \pm 0.6\text{ km s}^{-1}$, where the error is the epoch-to-epoch scatter, as a result of optically thick chromospheric self-absorption  (Fig.~\ref{fig:l3-chromo}). Collectively, these spectroscopic properties and the coronal plasma detected in soft X-rays by \textit{eROSITA} \citep{Freund:2024, Egbo:2026} are evidence for strong chromospheric magnetic heating through magnetic reconnection and wave dissipation, and they independently confirm the dwarf luminosity class of PM~J14111$-$6155.

To constrain the spectral type and fundamental stellar properties, we compared the flux-calibrated SALT/RSS spectrum against the empirical active M-dwarf templates of \citet{Bochanski2007} (see Fig.~\ref{fig:rss_templates}). When normalised to the pseudo-continuum between 5800 and 6250\,\AA, the active M1 template provides an excellent fit across the entire 3825--6302\,\AA\ baseline with a 4.4\% rms residual.
In contrast, the active M0 template yields a poor match ($\mathrm{rms} = 11.7\%$), systematically underpredicting flux by 20--40$\%$ blueward of 5000\,\AA. The M1 classification corresponds to an effective temperature of $T_{\mathrm{eff}} \approx 3600\text{--}3700$\,K \citep[e.g.][]{Pecaut2013}. Although the source was previously flagged in the \textit{Gaia}~DR3 Non-Single Star catalogue as an unresolved double main-sequence system \citep{Gaia2023_binaries}, the single active M1 template matches the optical continuum and molecular absorption bands down to $\sim$3800\,\AA\ without requiring a secondary component.

\section{Conclusions}

We have reported the serendipitous MeerKAT detection of a coherent radio burst from the high-proper-motion star PM\,J14111$-$6155, an M1Ve dwarf at a distance of $19.6$\,pc, identified in off-axis SMGPS observations. The burst is characterised by an extreme brightness temperature ($T_{\rm b}$ $> 10^{12}$\,K), $\sim100\%$ circular polarisation, an observed duration of $\gtrsim 20$\,minutes, and a restricted absolute bandwidth ($\Delta\nu \approx 300$\,MHz FWHM) exhibiting a slow negative frequency drift. These properties strongly favour ECMI originating in a local magnetic field of $B \approx 0.54\text{--}0.59$\,kG. These emission characteristics are closely analogous to the intrinsic auroral ECM emission observed in planetary magnetospheres and active low-mass stars.

Our supporting SALT observations provide definitive spectroscopic classification of the host star as a single M1 dwarf ($T_{\rm eff} \approx 3600\text{--}3700$\,K), demonstrating that the optical continuum matches standard empirical templates. High-resolution SALT/HRS spectra confirm vigorous chromospheric heating, evidenced by prominent \ion{Ca}{ii}~H\&K cores, core emission reversals in the \ion{Ca}{ii} infrared triplet, and double-peaked Balmer line emission. While optical monitoring with TESS across seven sectors between 2019 and 2026 reveals a moderate optical flare rate and no detected rotational modulation, such activity is typical for early-type M dwarfs. Ongoing high-precision spectroscopy with LCO/NRES (SAAO2025B-002) will place tighter limits on stellar radial-velocity variations and constrain rotation through $v \sin i$.

Beyond this specific active M dwarf, our results demonstrate the power of legacy MeerKAT synthesis imaging archives for uncovering medium-time-scale, highly polarised stellar transients. Systematic searches using dedicated pipelines such as TRON will transition these discoveries from serendipitous encounters to a robust statistical population of coherent stellar emission across the solar neighbourhood.

\paragraph*{Data Availability.} The raw SMGPS data underlying this work are publicly available through the SARAO archive\footnote{\url{https://archive.sarao.ac.za}}, under the proposal ID SSV-20180721-FC-01. The processed SALT spectra data presented are available in machine-readable format in the online supplementary material. Any additional data will be shared on reasonable request with the corresponding author.

\paragraph*{Acknowledgements.} The MeerKAT telescope is operated by the South African Radio Astronomy Observatory, which is a facility of the National Research Foundation, an agency of the Department of Science, Technology and Innovation. OMS's, JSK's and VGGS's research is supported by the South African Research Chairs Initiative of the NRF and DSTI (grant No. 81737). DAHB acknowledges research support from the National Research Foundation. OED is supported by an SAAO/SALT PhD Prize Scholarship. MAT gratefully acknowledges the support of the UK's Science \& Technology Facilities Council (STFC) through grant awards ST/R000905/1 and ST/W00125X/1. We acknowledge the financial support of the Breakthrough Listen project. Breakthrough Listen is managed by the Breakthrough Initiatives, sponsored by the Breakthrough Prize Foundation. We acknowledge the use of the ilifu cloud computing facility – www.ilifu.ac.za, a partnership between the University of Cape Town, the University of the Western Cape, Stellenbosch University, Sol Plaatje University and the Cape Peninsula University of Technology. The ilifu facility is supported by contributions from the Inter-University Institute for Data Intensive Astronomy (IDIA – a partnership between the University of Cape Town, the University of Pretoria and the University of the Western Cape), the Computational Biology division at UCT and the Data Intensive Research Initiative of South Africa (DIRISA). A part of this work is based on observations made with the Southern African Large Telescope (SALT) with the Large Science Programme on transients (2024-2-LSP-001; PI: DAHB). Polish participation in SALT is funded by grant No. MEiN 2021/WK/01.

\bibliographystyle{mnras} 
\bibliography{tron-mdwarf} 
 
\label{lastpage}

\end{document}